\documentstyle[preprint,eqsecnum,aps,epsf]{revtex}
\begin{document}

\hoffset-1cm

\draft
\preprint{CU-TH-97-05}

\title{
   Hadron Formation Time and Dilepton Mass Spectra\\
   in Heavy Ion Collisions
}

\author{Peter Filip and J\'an Pi\v s\'ut}

\address{
   Department of Theoretical Physics, Comenius University Bratislava,\\
   SK-842 15 Bratislava, Slovak Republic
}
%\date{\today}

\maketitle

\begin{abstract}
\baselineskip18pt
We point out that formation time of pions produced in heavy ion collisions
modifies the mass spectrum of dileptons produced via
$\pi ^+\pi ^- \longrightarrow e^+e^- $. Increasing formation
time enhances the production of dileptons with lower masses. The effect
offers an explanation of a part of the enhanced production of dileptons below
the $\rho $-meson mass as observed by the CERES and HELIOS
 Collaborations at the CERN SPS.
\end{abstract}

 %, 07.60.ly, 52.60.+h}

%%%%%%%%%%%%%%%%%%%%%%%%%%%%%%%%%%%%%%%%%%%%%%%%%%%%%%%%%%%%%%%%%%%%%%
\section{Introduction}\label{sec1}
%%%%%%%%%%%%%%%%%%%%%%%%%%%%%%%%%%%%%%%%%%%%%%%%%%%%%%%%%%%%%%%%%%%%%%

Dilepton production in heavy ion collisions is considered as a relatively
clear signature of the nature of matter produced in heavy-ion collisions.
This is one of reasons why the enhanced production of dileptons as
observed recently in heavy ion collisions by CERES \cite{c1,c2,c3},
HELIOS \cite{c4,c5} and NA-38 \cite{c6,c7} has received so much
of attention.

An attractive explanation of the peculiar dilepton mass shape is based
on the Brown-Rho \cite{c8} conjecture of decreasing $\rho $-meson mass
with increasing density and temperature of hadronic matter, related
to partial restoration of chiral symmetry.

When included into the transport codes the decreasing $\rho $-meson mass
together with effects due to higher nucleon and $\Delta $ resonancies
can lead to a qualitative agreement with data \cite{c9}. Good description
of dilepton mass spectra in S-Au and S-W interactions has been obtained
under the same assumption of the behaviour of $m_{\rho }$ by Li,Ko and
Brown \cite{c10}.

The purpose of the present note is to point out that the formation time
of pions produced in heavy-ion collisions leads also to modifications
of the shape of dilepton mass spectrum
 enhancing the production of lighter
dileptons with respect to heavier ones.

The concept of the formation time has appeared a long time ago
 in connection
with the problem of how fast is the electromagnetic field accompanying
a charged particle reconstructed after an abrupt change of the
momentum of the particle. Early literature on the question can be traced
back e.g. from Ref.\cite{c11}. In connection with production
of secondary particles in hadronic
collisions the concept has been introduced by Stodolsky \cite{c12}.
In analysis of data \cite{c13,c14} on multiparticle production in
proton - nucleus interactions it has turned out \cite{c13,c14,c15,c16}
that the cascading of secondaries is considerably lower than expected
under the assumption that a secondary pion is able to interact immediatelly
after it has been produced in a nucleon - nucleon collision and this
has been ascribed to the formation time of secondary particles.

The term "formation time" has somewhat different meaning in different
models. In what follows we shall understand this term as the time
interval between the time of the collision in which the secondary hadron
has been produced and the time when this hadron is able to interact
strongly (with the full cross-section) with other hadrons. For the
dilepton production in $\pi^+\pi^-$ interaction the formation time is
relevant, since the process goes via the intermediate $\rho $-meson
stage $\pi^+\pi^- \to \rho \to e^+e^- $ and the first part of the process
is strong.

We also assume that the formation time of a pion is
 characterized by $\tau _f$ which denotes the formation time of the pion
in its rest frame and in other frames it is Lorentz dilated.

To see in a very qualitative way the effect of formation time
on dilepton spectra consider a set of pions produced at $t=0$ in a volume
of typical dimension $L$. A pion with four-momentum $(E,\vec p)$ is unable
to interact before time $t=\tau _f \cdot E/m_{\pi}$ where $\gamma = E/m_{\pi}$
is the Lorentz time-dilation factor. Within this time the pion traverses
the distance $d=vt=\tau_f|\vec p|/m_{\pi}$. Faster pions traverse longer
distances before interacting. As a consequence thereof,
interactions of slower pions are prefered and this leads to softening
of dilepton spectra produced via $\pi ^+\pi ^- \longrightarrow e^+e^- $.

This mechanism will be important in situations when each of pions undergoes
only a few collisions. When the number of collisions, even for fast pions is
large, the system thermalizes and the spectrum of dileptons corresponds
to interactions of fully formed pions in such a thermalized system.

%%%%%%%%%%%%%%%%%%%%%%%%%%%%%%%%%%%%%%%%%%%%%%%%%%%%%%%%%%%%%%%%%%%%%%
\section{Scheme of calculation.}
%%%%%%%%%%%%%%%%%%%%%%%%%%%%%%%%%%%%%%%%%%%%%%%%%%%%%%%%%%%%%%%%%%%%%%

Suppose that the system of pions produced in a heavy ion collision is
described by a distribution function $f(k,x)$ in phase space,
where $x=(\vec x,t)$. The standard expression for the
rate of dilepton production is
(see e.g. \cite{c18}):

\begin{equation}
\frac{dN_{ee}}{dM^2d^4x} =
\int \frac{d^3k_1}{(2\pi)^3}f(k_1,x)
\int \frac{d^3k_2}{(2\pi)^3}f(k_2,x)
[v_r \sigma_e(M^2) \cdot \delta(M^2-(k_1+k_2)^2)]
\label{1}
\end{equation}

where $\sigma_e(M)$ is the cross-section for
$\pi ^+\pi ^- \longrightarrow e^+e^- $ at given $\sqrt{s}=M$,
$s=(k_1+k_2)^2$, and $v_r$ is the relative velocity.
 Distribution functions $f(k,x)$ depend on the cross-section
$\sigma_{\pi}$ for  $\pi ^+\pi ^- \longrightarrow \pi ^+\pi ^- $
which regulates scattering of pions, and on the
 value of the formation time.
Multiplying and dividing the right-hand side of Eq.(2.1) by
$\sigma _{\pi}$ and integrating over $d^4x$ we obtain

\begin{equation}
\frac{dN_{ee}}{dM^2} =
R\cdot \frac{\sigma_e(M^2)}{\sigma_{\pi}(M^2)}
\int d^4x \frac{d^3k_1}{(2\pi)^3}f(k_1,x)
\frac{d^3k_2}{(2\pi)^3}f(k_2,x) [\delta(M^2-(k_1+k_2)^2) \cdot
v_r \sigma_{\pi}(M^2)]
\label{2}
\end{equation}

which can be rewritten as
\begin{equation}
\frac{dN_{ee}}{dM^2} =
R\cdot \frac{\sigma_e(M^2)}{\sigma_{\pi}(M^2)}
\cdot \frac{dN_{\pi}}{dM^2}
\label{3}
\end{equation}

where $\frac{dN_{\pi}}{dM^2}$ is simply
distribution of the number of $\pi\pi$ collisions
as a function of the mass of the $\pi\pi$ system and $R$ denotes
the ratio of the number of $\pi ^+\pi ^-$ collisions to all
$\pi \pi $ interactions.
The expression
$\frac{dN_{\pi}}{dM^2}$ is calculated by Monte-Carlo method in a given
cascade code. The term $\sigma_{\pi}(M^2)$ is the total $\pi\pi$ cross-section
including $s-$ and $p-$ waves plus possibly higher ones.
The term $\sigma_e(M^2)$ is the $\pi^+\pi^-\longrightarrow e^+e^-$
cross-section which can be taken in the simplest approximation as

\begin{equation}
\sigma _e=\frac{4\pi}{3} \frac{\alpha ^2}{s}
\Big[1-\frac{4m_{\pi}^2}{s}\Big]^{\frac{1}{2}}
\Big[1-\frac{4m_{e}^2}{s}\Big]^{\frac{1}{2}}
\Big[1+2\frac{m_e^2}{s}\Big]|F_{\pi}|^2
\label{4}
\end{equation}
This result follows directly from a calculation of the Feynman diagram.
Neglecting the electron mass $m_e$ with respect to $\sqrt{s}=M$ we
obtain

\begin{equation}
\sigma _e=\frac{4\pi}{3} \frac{\alpha ^2}{s}
\Big[1-\frac{4m_{\pi}^2}{s}\Big]^{\frac{1}{2}}
|F_{\pi}|^2
\label{5}
\end{equation}

The form- factor squared $|F_{\pi}|^2$ can be simply parametrized
in the Bright-Wigner approximation as

\begin{equation}
|F_{\pi}|^2 =
\frac{m_{\rho }^4 + m_{\rho }^2\Gamma_{\rho }^2}
     {(M^2-m_{\rho}^2)^2+m_{\rho}^2\Gamma_{\rho}^2}
\label{6}
\end{equation}

In practical calculations we have used a multiple scattering model \cite{c19}
inspired by that used by Humanic \cite{c20}, see also Ref.\cite{c21}.
Since $\pi\pi$ interactions are most likely responsible for the enhanced
production of dileptons with masses below 1GeV/$c^2$ we have studied only
a simple version of the multiple scattering model in which all secondary
hadrons are pions. The initial state for the evolution of the cascade is
given by set of momenta and positions in which pions are created in
nucleon - nucleon sub-collisions. This initial state has been generated
by the Monte-Carlo model of Z\'avada \cite{c16}.
We have used isospin
averaged $\pi\pi$ cross section (see e.g. Appendix in Ref.\cite{c22}):

\begin{equation}
{\frac {d\sigma_{\pi}(s,\theta)}{d\Omega}} =
{\frac {4}{q^2}}
\left(
 {1\over 9}sin^2\delta^0_0+{5\over 9} sin^2\delta^2_0+
 {27\over 9}sin^2\delta^1_1 cos^2\theta
\right)
\label{7}
\end{equation}

 The total cross-section is obtained by integrating over a half of the
angular space (due to identity of particles)
\begin{equation}
\sigma_{\pi}(s)={\frac {8\pi}{q^2}}\left( {1\over 9}sin^2\delta^0_0
+{5\over 9}sin^2\delta^2_0 + sin^2\delta^1_1 \right)
\label{8}
\end{equation}

Here $\delta^T_l$ is the phase shift in the partial wave with isospin $T$
and orbital momentum $l$, $\theta$ is the CMS scattering angle, $s$ is the CMS
energy squared and $q=(1/2)\sqrt{s-4m^2_{\pi}} $ denotes the CMS momentum.

Rescattering program follows trajectory of each pion in short time
steps and when the distance between two pions is smaller than
$\sqrt {\sigma_{\pi}(s)/\pi}$ the pions scatter according the cross-
section given by Eq.(2.7). The scattering angle is determined in CMS frame
of two pions being scattered and then momenta of pions are
transformed back to the global frame of the simulation.
 The scattering is assumed \cite{c19,c20} to
take time interval $t_i$ of the order of 1fm/c, which roughly corresponds
to $\hbar/\Gamma_{\rho}$. During this time interval the participating
pions are unable to interact again.

In contradistinction to other Monte-Carlo cascade codes,
like e.g.\cite{c20} we take initial conditions directly from a realistic
model \cite{c16} of pion production in nuclear collisions and
 introduce the formation time $\tau_f$ in the way discussed in Ref.\cite{c24}.
The formation time characterizes the formation of pion in its rest frame.

  For times lower than Lorentz dilated formation time
  $(E_{\pi}/m_{\pi})\tau_f$
 a pion created in the initial state is unable to interact.
The formation time is Lorentz dilated, whereas the time $t_i$ is fixed in the
global frame of simulation which is CMS of Pb+Pb system.
Rate of collisions per pion varies from  3.7 for $\tau _f=0.2 fm$ to
1.0 for $\tau = 1.0 fm$ in our simulation. We guess that
effects of non-locality (see e.g. \cite{c29})
present also in our rescattering simulation do not influence substantially
results for dilepton mass spectrum.

\section{Results of the simulation}
We have run the rescattering program for three different values
of formation time parameter $\tau _f$ using initial state of pion gas
generated by cascading generator \cite{c16} for
Pb+Pb 160 GeV/n b=7fm collisions.
Mass spectrum of dileptons was calculated according to the equation (2.3)
from the mass spectrum of $\pi\pi$ collisions obtained from the simulation.
Cross sections given by (2.4) and (2.8) were properly normalized and resulting
distribution of dileptons was rescaled to the $100$ MeV bin size,
charged multiplicity $dN_{ch}/d\eta = 300$ and yield of dileptons per
unit of rapidity. In Fig.1 we present the shape and magnitude of dilepton
mass spectrum as a function of the formation time $\tau _f$. The
interpretation of this dependence is simple. Dilepton yields
for low masses are almost independent on $\tau _f$ since slower pions
are formed rather fast. On the other hand dileptons with masses in the
$\rho $-meson region are suppressed for larger values of the formation time.
This is due to the Lorentz dilation factor $(E_{\pi }/m_{\pi})$ which
permits faster pions to escape from the interaction region with little
or no rescattering or annihilation.

\vskip0.7cm
\centerline{\epsfxsize=10.7cm\epsffile{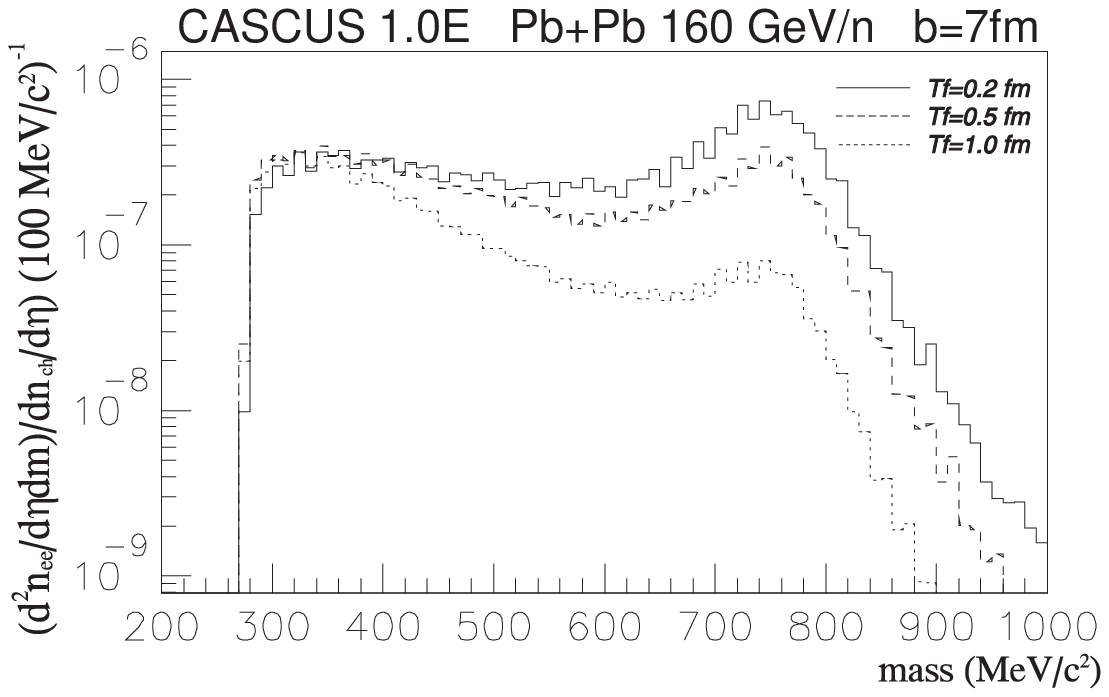}}
\vskip1.7pt
\centerline{\parbox{13cm} {\small {\bf Fig.1} Dilepton
invariant mass distributions obtained by rescattering simulation
for three different values of formation time parameter.
}}
\vskip0.5cm

In Fig.2 we compare the excess of dileptons as observed by CERES
collaboration \cite{c25} with our calculations of dilepton production
due to $\pi \pi $ annihilation as given in Fig.1.
Data points and the line which gives the expected dilepton
production as extrapolated from proton-nucleus interactions are
taken from \cite{c25}. Our results
presented in Fig.1 are added to the extrapolated continuous line.

\vskip0.7cm
\centerline{\epsfxsize=10.7cm\epsffile{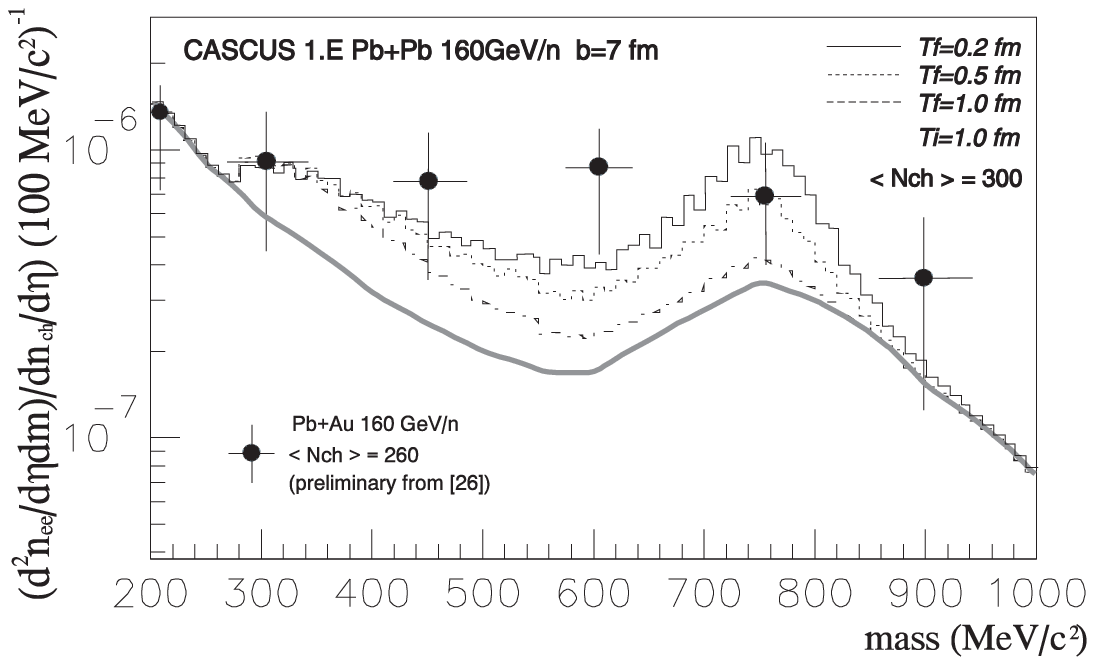}}
\vskip1.7pt
\centerline{\parbox{13cm} {\small {\bf Fig.2} Excess of dileptons
from $\pi\pi$ annihilation in hadron gas. Data points and
background are taken from
\cite{c25}.
}}
\vskip0.5cm

The comparison shows that a rough and qualitative agreement could be
reached for formation times $\tau _f \approx 0.5$ fm. This comparison
(Fig.2) can only serve as a very qualitative argument,
since the extrapolation
procedure usually used to get the background line in Fig.2 is based on
assumption that $\rho $-meson dilepton decays and other contributions
to dilepton production can be simply rescaled from proton-nucleus
to nucleus-nucleus collisions. This may be quite a rough approximation
since $\rho $-mesons and other resonancies rescatter in Pb-Pb collisions
with much larger probability than in proton-nucleon case.
We think that quantitative analysis can be based only on a detailed
Monte-Carlo code containing $\rho $-mesons and other resonancies
in the initial state and taking their rescattering into account.

\section{Comments and conclusions}
We have shown above that the value of formation time $\tau _f$
strongly influences the shape and magnitude of the mass
spectrum of dileptons produced in $\pi \pi $ annihilation.
This concerns in particular the ratio of low mass dileptons to those
in the $\rho $-meson region.

We have also shown that the shape and order of magnitude of the dilepton
production due to $\pi ^+ \pi ^- $ annihilations roughly corresponds
to the excess of dileptons as observed by the CERES and HELIOS collaborations
at the CERN SPS, both in the low mass and $\rho $-meson region.
Contribution around $m_{ee} \approx 500-600 $ MeV/c$^2$ may still be
missing.

The excess of dileptons as observed at CERN-SPS is of a similar
shape as that observed by DLS collaboration \cite{c26}. We think that
the  question whether formation time effects in calculations
like \cite{c27,c28} can bring models closer to data
deserves a detailed study.

\vskip2cm
{\bf Aknowledgements:} We are indebted to Peter Z\'avada for generating
the initial state for our rescattering code and to A.Nogov\'a for
collaboration at the early stages of this study.
This work was supported by Grant Committee VEGA G-F01 in Slovak Republic.

%%%%%%%%%%%%%%%%%%%%%%%%%%%%%%%%%%%%%%%%%%%%%%%%%%%%%%%%%%%%%%%%%%%%%%

%%%%%%%%%%%%%%%%%%%%%%%%%%%%%%%%%%%%%%%%%%%%%%%%%%%%%%%%%%%%%%%%%%%%%%

\begin{thebibliography}{99}
\bibitem{c1}
  G.Agakichiev et al., Phys. Rev. Lett. {\bf 75}
  (1995) 1272

\bibitem{c2}
  A.Drees, Nucl.Phys A {\bf 610}
  (1996) 536c

\bibitem{c3}
  A.Drees, QCD Phase Transitions, Proc. of the XXV Int. Workshop, Hirschegg,
  Jan. 13-18, 1997, p.178, Ed. H.Feldmeier et al., GSI Darmstadt 1977

\bibitem{c4}
  I.Kr\'alik, PhD Thesis, Slovak Academy of Sciences, Ko\v sice 1995

\bibitem{c5}
  M.Masera et al., Nuclear Physics A {\bf 590} (1995) 93c

\bibitem{c6}
  C.Lourenco, Proc. of the XXIII Int.Workshop on Gross Properties
  of Nuclei and Nuclear Excitations, Hirschegg, 1995

\bibitem{c7}
  C.Lourenco, Doctoral Thesis, Universidade Tecnica de Lisboa (1995)

\bibitem{c8}
  G.E.Brown and M.Rho, Phys.Rev.Lett. {\bf 66}
  (1991) 2720

\bibitem{c9}
  W.Cassing et al., Phys.Lett. B {\bf 363}
  (1996) 35; U.Mosel see Ref.[3] page 201; W.Cassing et al.,
  Phys.Lett. B {\bf 377} (1996) 5; B.Friman, Nucl.Phys. {\bf A610}
  (1996) 358c

\bibitem{c10}
  G.Q.Li, C.M.Ko and G.Brown, Phys.Rev.Lett. {\bf 75} (1995) 4007

\bibitem{c11}
  E.L.Feinberg, p.248 in Problems of Theoretical Physics, Memorial
  volume to I.E.Tamm, Nauka, Moscow, 1972, Ed. V.L.Ginzburg et al.;
  N.N.Nikolaev, Soviet Physics, Uspekhi {\bf 24} (1981) 531

\bibitem{c12}
  L.Stodolsky, Phys.Rev.Letters {\bf 28} (1972) 60

\bibitem{c13}
  C.De Marzo et al., Phys.Rev. {\bf D 26} (1982) 1019; Phys.Rev. {\bf D 29}
  (1984) 363 and Phys.Rev. {\bf D 29} (1984) 2476

\bibitem{c14}
  D.A.Brick et al., Phys.Rev. {\bf D 39} (1989) 2484

\bibitem{c15}
  J.Ranft, Phys.Rev. {\bf D 37} (1988) 1842; Zeit.f.Phys. {\bf C 43}
  (1989) 439 and H.J.M\"{o}hring and J.Ranft, Zeit.f.Phys. {\bf C 52}
  (1991) 643

\bibitem{c16}
  P.Z\'avada, Zeit.f.Phys. {\bf C 32} (1986) 135; Phys.Rev. {\bf C 40}
  (1989) 285 and Phys.Rev. {\bf C 42} (1990) 1104

\bibitem{c17}
  B.B.Levchenko and N.N.Nikolaev, Yad.Fiz. {\bf 37} (1983) 1016
  and Yad.Fiz. {\bf 42} (1984) 1255

\bibitem{c18}
  P.V.Ruuskanen, Dilepton emission in reletivistic nuclear collisions
 in {\it Quark- Gluon Plasma 1}, edited by R.C. Hwa, World Scientific,
 1993, Singapore

\bibitem{c19}
  P.Filip, Acta Physica Slovaca {\bf 46} (1996) 9; {\bf 47} (1997) 53

\bibitem{c20}
  T.Humanic, Phys.Rev. {\bf C 50} (1994) 2525; Phys. Rev.{\bf C53}
  (1996) 901

\bibitem{c21}
  M.Hermann and G.F.Bertsch, Phys. Rev. {\bf C 51} (1995) 328

\bibitem{c22}
M.Prakash et al., Phys. Reports {\bf 227} (1991) 321

\bibitem{c23} C.D.Froggatt and J.L.Petersen,
	     Nucl.Phys. {\bf B129} (1977) 89

\bibitem{c24} J.Pi\v s\'ut, N.Pi\v s\'utov\'a and P.Z\'avada, Zeit. f. Phys.
       {\bf C67} (1995) 467

\bibitem{c29} G.Kortemeyer et al., Phys.Rev. {\bf C52} (1995) 2714;
 P.Danielewicz and S.Pratt, Phys.Rev. {\bf C53} (1996) 249


\bibitem{c25} G.Agakichiev et al., Nucl. Phys. {\bf A610} (1996) 317c

\bibitem{c26} G.Roche et al., DLS Collaboration, Phys.Rev.Lett. {\bf 61}
(1988) 1069; H.Z.Huang et al., DLS Coll., Phys. Rev. {\bf C49}
(1994) 314; G.Roche, Acta Phys. Slov. {\bf 44} (1994) 127

\bibitem{c27} J.Kapusta and P.Lichard, Phys.Rev. {\bf C40} (1989) R1574;
K.Haglin and C.Gale, Phys.Rev. {\bf C49} (1994) 401

\bibitem{c28} L.A.Winckelmann et al., Phys.Letters {\bf B298} (1993) 22;
L.A.Winckelmann et al., Phys.Rev. {\bf C51} (1995) R9

\end{thebibliography}
\end{document}